\documentclass[aip,apl,preprint,graphicx]{revtex4} 
\usepackage{graphicx}
\usepackage{bm}

\usepackage{epstopdf}
\usepackage{dcolumn}

\usepackage{gensymb}

\begin{document}


\title{A planar Al-Si Schottky Barrier MOSFET operated at cryogenic temperatures}

\author{W. E. Purches}
\affiliation{School of Physics, UNSW, Sydney, 2052, Australia}
\author{A. Rossi}
\affiliation{School of Electrical Engineering and Telecommunications, UNSW, Sydney, 2052, Australia}
\author{R. Zhao}
\affiliation{School of Electrical Engineering and Telecommunications, UNSW, Sydney, 2052, Australia}
\author{S. Kafanov}
\affiliation{School of Physics, UNSW, Sydney, 2052, Australia}
\affiliation{Centre for Engineered Quantum Systems (EQuS), School of Physics, UNSW, Sydney, 2052, Australia}
\author{T. L. Duty}
\affiliation{School of Physics, UNSW, Sydney, 2052, Australia}
\affiliation{Centre for Engineered Quantum Systems (EQuS), School of Physics, UNSW, Sydney, 2052, Australia}
\author{A. S. Dzurak}
\affiliation{Australian Centre of Excellence for Quantum Computation and Communication Technology (CQC$^{2}$T), UNSW, Sydney, 2052, Australia}
\affiliation{School of Electrical Engineering and Telecommunications, UNSW, Sydney, 2052, Australia}
\author{S. Rogge}
\affiliation{School of Physics, UNSW, Sydney, 2052, Australia}
\affiliation{Australian Centre of Excellence for Quantum Computation and Communication Technology (CQC$^{2}$T), UNSW, Sydney, 2052, Australia}
\author{G. C. Tettamanzi}
\email{g.tettamanzi@unsw.edu.au}
\affiliation{School of Physics, UNSW, Sydney, 2052, Australia}
\affiliation{Australian Centre of Excellence for Quantum Computation and Communication Technology (CQC$^{2}$T), UNSW, Sydney, 2052, Australia}

\date{\today}

\begin{abstract} Schottky Barrier (SB)-MOSFET technology offers intriguing possibilities for cryogenic nano-scale devices, such as Si quantum devices and superconducting devices. We present experimental results on a device architecture where the gate electrode is self-aligned with the device channel and overlaps the source and drain electrodes. This facilitates a sub-5~nm gap between the source/drain and channel, and no spacers are required. At cryogenic temperatures, such devices function as p-MOS Tunnel FETs, as determined by the Schottky barrier at the Al-Si interface, and as a further advantage, fabrication processes are compatible with both CMOS and superconducting logic technology. 
\end{abstract}

 \maketitle

Schottky Barrier Metal Oxide Semiconductor Field Effect Transistors (SB-MOSFETs) offer multiple advantages over typical Complementary Metal Oxide Semiconductor (CMOS) device structures, particularly in the area of scalability~\cite{Larson}, however, they suffer a major challenge imposed by the requirement for a low Schottky barrier ($<$~0.15~eV) in order to exceed the performance of CMOS devices at room temperature~\cite{Connelly}. For this reason their uptake has been limited, but interest remains in areas such as low power biomedical applications~\cite{Cabral}, bio-sensing~\cite{Zorg} and in particular, nano-devices~\cite{Choi}. In this letter we discuss and investigate advantages which arise at cryogenic temperature.

Recent advances in Si-based quantum electronics~\cite{Rogge} demand development of components optimized for cryogenic operation in order to complement quantum components at the Si on-chip integration level. SB-MOSFET technology holds strong potential in this area because at cryogenic temperature ($<$~77~K) device carrier transport is dominated by Fowler-Nordheim tunneling, enabling Tunnel FET functionality~\cite{Lepselter}. This allows insensitivity to thermal fluctuations, enabling low-power/low-noise on-chip amplification for cryogenic Si quantum devices~\cite{Sedighi} or nano-scale switches integrated into superconducting devices, with the Schottky barrier preventing proximity effect~\cite{Satoh}. Another potential application is replacement of spacers used to provide potential barriers in cryogenic nano-devices, which can act as unwanted sources of capacitance~\cite{Urda}.

All advantages offered at room temperature, are maintained at cryogenic temperature, such as reduced source/drain parasitic resistance, reduction of the off-state leakage current between source and drain, elimination of parasitic bipolar action, a lower thermal budget and fewer processing steps~\cite{Larson}. Furthermore, problems with off-state thermionic gate leakage induced by work function differences between gate electrode and source/drain electrodes are eliminated \cite{Snyder}.

\begin{figure} 
\begin{center}
\includegraphics[bb=0 0 244 151]{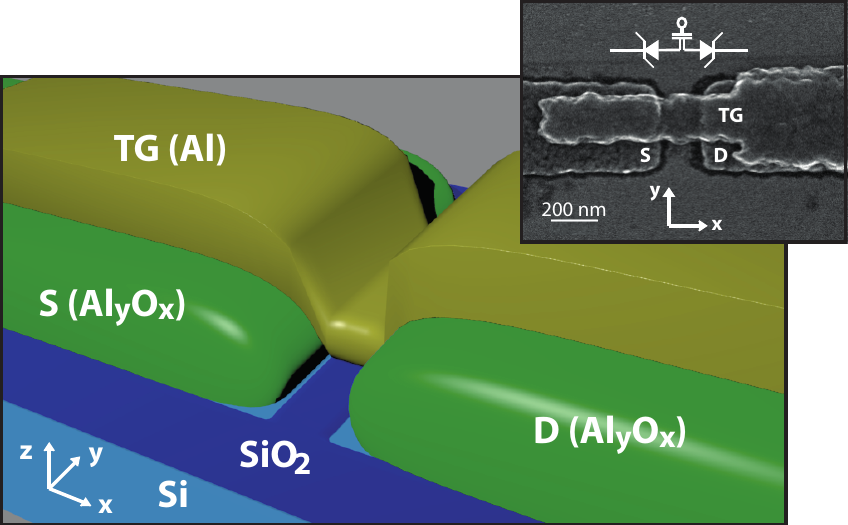}
\caption{Schematic illustration of device architecture. The Al source (S) and drain (D) regions are colored bright green, showing the Al$_{y}$O$_{x}$ thermally grown on the surface, with Al top gate (TG) colored light green. Regions of SiO$_{2}$ which may have been over etched are shown near edge of source and drain in the light blue of the underlying Si. Inset: Scanning Electron Microscopy (SEM) image of Device 1 and equivalent circuit diagram.}
\label{fg:pmos}
\end{center}
\end{figure}

In this work we report the fabrication, simulation and electrical characterization of an SB-MOSFET device at a fridge temperature as low as 10~mK. The self-aligned gate-electrode overlaps the source and drain and negates any requirement for side-wall spacers, differing from the typical ``co-planar'' Si-based design offered by self-aligning silicides~\cite{Moon}. Our design, shown in Fig.~\ref{fg:pmos}, minimizes spacing between the device channel and Schottky barrier, also known as ``underlap'', which increases drive current~\cite{Snyder}.

Two devices of different channel dimensions were investigated; Device 1, of nominal channel length 175~nm and width 245~nm (Fig.~\ref{fg:pmos}), and Device 2, of nominal channel length 90~nm and width 275~nm. They were fabricated on a high-purity near-intrinsic 200~$\mu$m thick Si substrate with a 5~nm thick thermally grown layer of SiO$_{2}$ on the surface acting as gate dielectric. A two-layer Al stack was defined by using electron-beam lithography and thermal evaporation~\cite{Angus, Jove}.

\begin{figure*} 
\begin{center}
\includegraphics[bb=0 0 487 264]{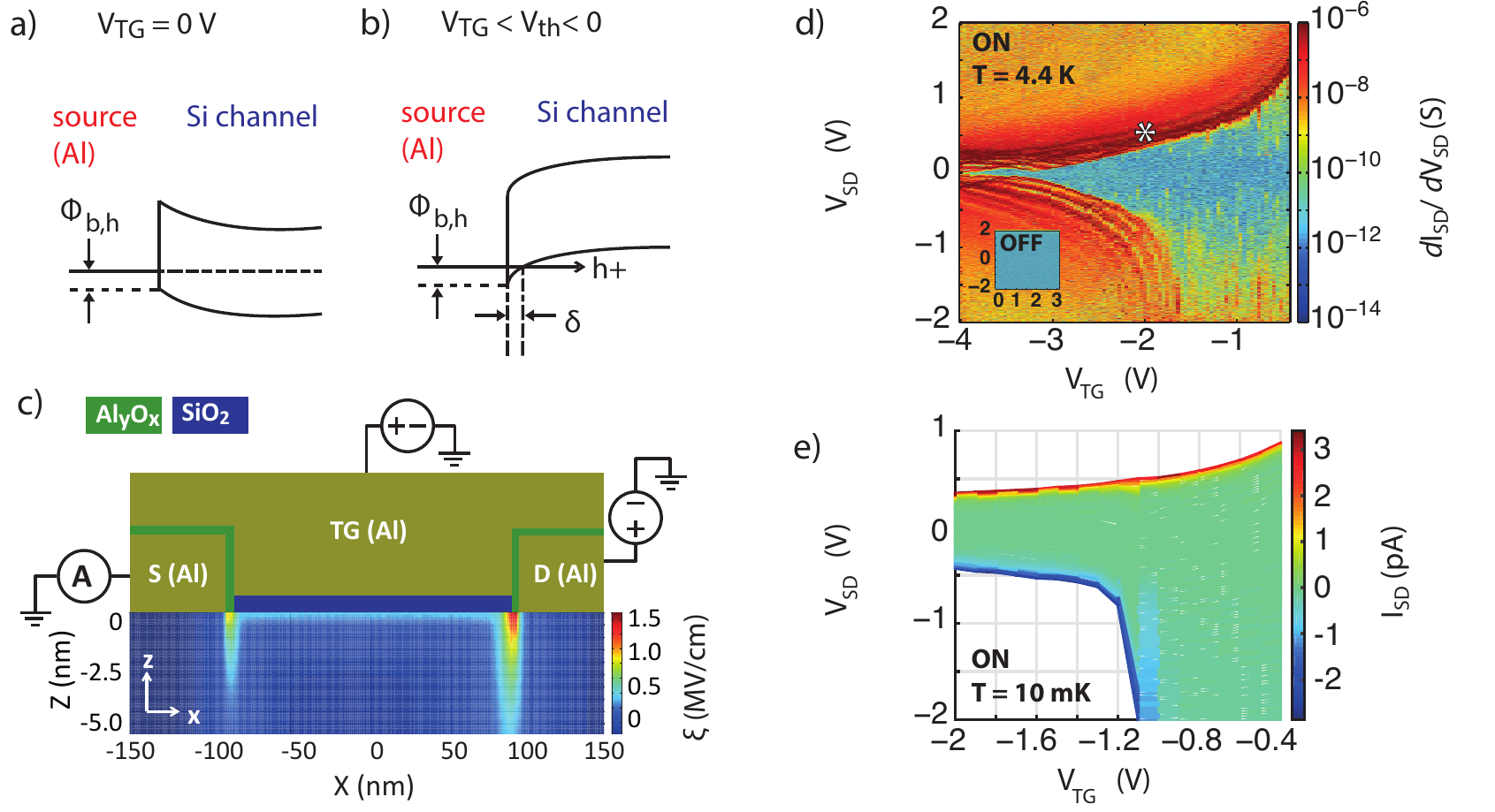}
\caption{a) Energy band diagram for device switched OFF; neither holes or electrons can enter the channel. b) Energy band diagram for device switched ON; holes can enter the channel via tunneling. c) Synopsis TCAD simulation of the absolute electric field for device while ON, with V$_{TG}$~=~-2~V and V$_{SD}$~=~+0.5~V. d) Experimental measurement of device 1 at T~=~4.4~K, differential conductance as a function of negative V$_{TG}$ and V$_{SD}$. White asterisk shows position of shown TCAD simulation. Inset: Data for positive V$_{TG}$'s for same device and temperature. e) Experimental measurement of device 2 at T~=~10~mK, current as a function of negative V$_{TG}$ and V$_{SD}$. Current in white regions is beyond limits of equipment set-up ($>$~5~pA).}
\label{fg:pmos2}
\end{center}
\end{figure*}

The bottom layer formed the Al/Si Schottky source (S) and drain (D) electrodes. By selectively removing SiO$_{2}$ from lithographically defined regions of the chip by immersion in hydrofluoric acid, nano-scale metal-semiconductor junctions were realized. This anisotropic wet-etch can lead to over-etching in the planar direction in excess of lithographically defined feature sizes (Fig.~\ref{fg:pmos}), therefore extremely careful calibration of etching time and rate was required. The time between etching and metal deposition was minimized in order to avoid the growth of native SiO$_{2}$, which would prevent the formation of metal-semiconductor junctions.

The top Al layer was used to define the gate electrode to control the density of charge carriers in the semiconductor channel between source and drain. As shown in Fig.~\ref{fg:pmos}, the top gate partially overlaps the bottom metal electrodes. Inter-layer electrical insulation was ensured by thermal oxidation of the bottom Al layer prior to deposition of the top gate~\cite{Angus}. The few~nm oxidation thickness determines the ``underlap'' of the device. Finally a 30~min thermal anneal at 400~\degree~C was carried out in forming gas to improve the quality of the contacts. The Al S/D leads, partially shown in Fig.~\ref{fg:pmos}, are few hundreds of $\mu$m in length and are terminated with rectangular regions of about 200x50~$\mu$m$^2$ used as bond pads. These S/D pads are then used to contact the devices, via Al bonding wires, to a printed circuit board ultimately connecting them to the room temperature electronics as schematically shown in Fig.~\ref{fg:pmos2}c).

Measurements on Device 1 were carried out in a liquid helium cryostat with a Heliox ($^{3}$He refrigeration) variable temperature controller from Oxford Instruments, in the temperature range 4.4~K to 50~K, while electrical measurements were taken with a low noise current-voltage measurement system~\cite{Koppens}. Measurements on Device 2 were performed in a BlueFors dilution refrigerator of base temperature 10~mK in the temperature range 10~mK to 1.6~K with RC filtering on all DC lines, while electrical measurements were taken with a Stanford Research Isolated Voltage Source (SIM 928) and Keithley 6430 Sub-Femtoamp ammeter.

In a SB-MOSFET, transport is principally determined by the Schottky barrier height between the source/drain electrodes and the Si. An equivalent circuit diagram is shown in the inset of Fig.~\ref{fg:pmos}, illustrating that when the device is in operation, current will effectively flow through two Schottky diodes in series, one of which is in reverse bias, while the other is in forward bias~\cite{Sze4}. Depending on the source/drain bias polarity, one contact will operate in reverse mode, while the other one will be in the active mode. As a consequence, the transport of the device will be dominated by the reverse biased contact. Hence, at cryogenic temperatures ($<$~77~K), the current is dominated by tunneling, as carriers lack the thermal energy to overcome the reverse biased Schottky barrier~\cite{Lepselter}. Tunneling current is determined by the width of the tunneling barrier ($\delta$) at the metal/Si Schottky interface, which is controlled by both the height of the Schottky barrier and the electric field in the Si, induced by both the top gate voltage (V$_{TG}$) and the source/drain voltage (V$_{SD}$)~\cite{Sze2}.

An Al-Si Schottky contact formed in our device processing conditions, will have an expected barrier height of around 0.35~eV for holes ($\phi_{b,h}$) and 0.75~eV for electrons at room temperature~\cite{Card}, thus favoring p-MOS operation~\cite{Larson}. For this case, at cryogenic temperature, when V$_{TG}$~=~0~V, neither thermionic or tunneling charge transport across the Schottky barrier is possible. This is illustrated in Fig.~\ref{fg:pmos2}~a). As the V$_{TG}$ reaches the negative threshold voltage (V$_{th}$), the electric field in the silicon creates band-bending at the Schottky contact, such that the tunneling barrier at the reverse biased Schottky contact becomes narrow enough in width to enable hole transport, and the device switches on. This is illustrated in Fig.~\ref{fg:pmos2}~b). For a barrier height of 0.75~eV for electrons, the tunneling barrier may only be narrowed enough to enable electron transport under exceptionally high electric fields~\cite{Tucker}.

\begin{figure} 
\begin{center}
\includegraphics[bb=0 0 244 193]{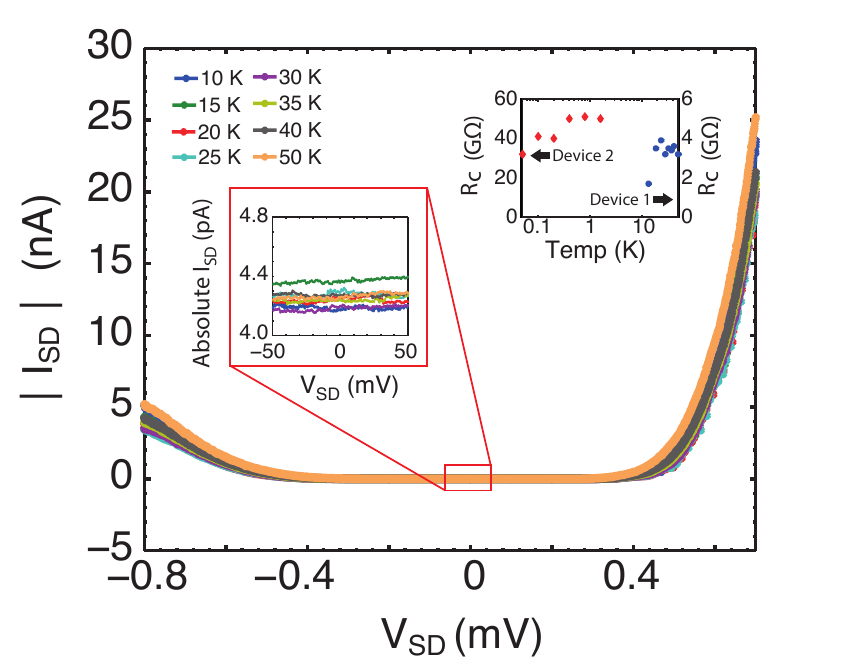}
\caption{IV plots across the temperature range 10~-~50~K for V$_{TG}$~=~-~2~V, demonstrating negligible change with temperature.  Left inset shows the region around V$_{SD}$ = 0, that was used to calculate R$_{C}$. Right inset shows R$_{C}$ as a function of temperature for both Device 1 and 2.}
\label{fg:TEMP}
\end{center}
\end{figure}
In order to further understand charge transport in our device, the electric field of Device 1 was simulated using the commercial software Synopsis TCAD~\cite{Rossi} for a range of experimental bias voltages, with an example cross section of a simulation shown in Fig.~\ref{fg:pmos2}~c). It was observed that the relevant field strengths in the Si were most significant in both the z and x direction; in the x-direction, directing holes from the source (V$_{SD}$~$<$~0) or drain (V$_{SD}$~$>$~0) into the channel region, and in the z direction, directing holes from the source or drain into the Si. Hence the absolute magnitude of electric field in the Si ($\xi$) is shown in Fig.~\ref{fg:pmos2}~c) for conditions where the device is expected to be ON (V$_{TG}$~=~-2~V and V$_{SD}$~=~+0.5~V), as shown in Fig.~\ref{fg:pmos2}~b), with tunnel current flow enabled at the reverse biased drain contact. This was shown to be consistent with electrical measurements of Device 1 and 2.

For all electrical measurements, bias at the drain contact and gate electrode was varied while source was grounded at all times, as shown in Fig.~\ref{fg:pmos2}~c). It was confirmed that both Device 1 and 2 operated as p-MOS devices in the temperature range 10~mK to 4.4~K, switching on with a negative gate bias, as shown in Fig.~\ref{fg:pmos2}~d) and e). Inset of Fig.~\ref{fg:pmos2}~d) shows how device does not switch on for range of positive V$_{TG}$ tested. In Fig.~\ref{fg:pmos2}~d) an asterisk marks the conditions for the simulation results shown in Fig.~\ref{fg:pmos2}~c). It can be noted that there is asymmetrical current behavior around V$_{SD}$~=~0 for both Device 1 and 2. This is because the dominant, reverse biased electrode, experiences an electric field which is heavily influenced by the difference between its reference voltage and the top gate voltage. Given that the source is always biased 0~V in reference to ground, with only the drain bias adjusted during measurement, as shown in Fig.~\ref{fg:pmos2}~c), the electric field at the drain is always stronger than the electric field at the source. Therefore the reverse biased drain will always actuate a higher current for any given magnitude of V$_{SD}$, than that determined by the reverse biased source at the same magnitude of V$_{SD}$.

The mechanism of charge transport was investigated via I-V curves in the temperature range 10~-~50~K for Device 1 and 50~-~1600~mK for Device 2; in both cases at a gate voltage of V$_{TG}$~=~-2~V, with data shown for Device 1 in Fig.~\ref{fg:TEMP}. The gate voltage was chosen to ensure the device was on. Contact resistance ($R_{C}$) was calculated from these plots (left inset shows data points used) at each temperature (right inset) and for both devices by using Eq.~\ref{eq:Rc}~\cite{Sze2}.

\begin{equation}
R_{C}~=~\Big( \frac{\partial I_{SD} }{\partial V_{SD}} \Big)^{-1}_{V_{SD}=0}
\label{eq:Rc}
\end{equation}

\noindent Where $I_{SD}$ is the current. $R_{C}$ is almost constant as a function of temperature, which is a strong indicator that tunneling is the dominant form of carrier transport, as R$_{C}$ decreases with increasing temperature if dominated by thermionic emission~\cite{Tettamanzi}, and remains constant if dominated by tunneling~\cite{Chang}. It can be noted that the R$_{C}$ of Device 2 is larger than Device 1. It is probable that this is due to the effect of device processing conditions on the Al-Si interface. Similar devices having Al-Si contacts on degenerately doped n-type Si reported in the literature achieved much lower R$_{C}$s than the ones of our devices~\cite{Muhonen}, however, this is to be expected as Si doping greatly reduces contact resistance. Furthermore, the step in the $R_{C}$ value observed for Device 2 at $T$~$\approx$~1~K might be related to the superconducting/normal transition in the Al leads~\cite{Tinkham}. It is possible to describe the tunneling current of our devices by Eq.~\ref{eq:J}~\cite{Zhang}.

\begin{equation}
J_{SD} = \frac{q^{3}\xi V_{SD}}{4 \pi^{2}\hbar^{2}} \sqrt{\frac{m_{l,h}}{2\phi_{b,h}}}exp \Big(-\frac{4\sqrt{2m_{l,h}}\phi_{b,h}^{3/2}}{3q\xi\hbar} \Big)
\label{eq:J}
\end{equation}

\begin{figure} 
\begin{center}
\includegraphics[bb=0 0 244 277]{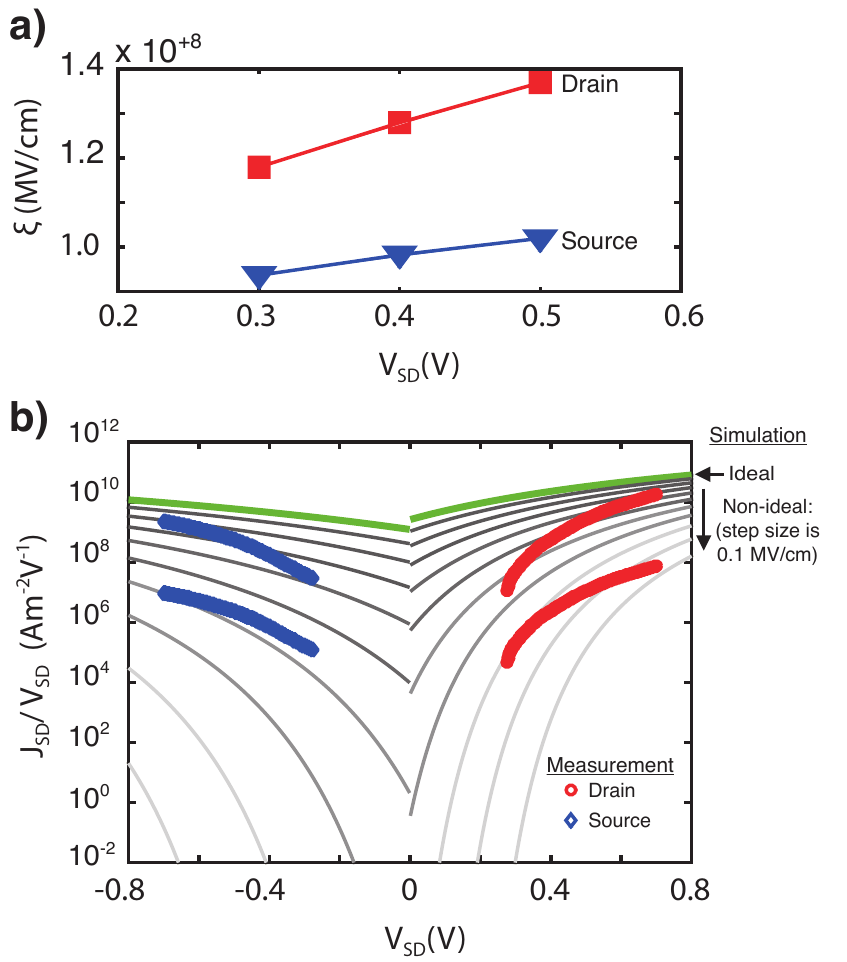}
\caption{a) Synopsis TCAD results for various experimental V$_{SD}$. b) J$_{SD}$ normalised against V$_{SD}$ vs V$_{SD}$ for measured and simulated results. Calculated upper and lower limits of normalised J$_{SD}$ for source and drain I$_{SD}$ measurements are shown. Green bold lines show the simulated ideal case (no offset). Grey lines show the non-ideal cases affected by non-abruptness at the Al-Si interface, with a step size in introduced $\xi$ offset of 0.1~MV/cm.}
\label{fg:MOD}
\end{center}
\end{figure}

\noindent Where $J_{SD}$ is current density, $q$ is the charge on an electron, $\hbar$ is the reduced Planck constant and $m_{l,h}$ the effective hole mass for light holes (taken as a ratio of 0.16 of the electron mass~\cite{Shin,Calvet2000}). $\xi$ was estimated by Synopsis TCAD simulations for V$_{TG}~=~-2~V$ and six experimental values of V$_{SD}$ as shown in Fig.~\ref{fg:MOD}~a), and taken as the maximum field strength at a position beneath the edge of the drain contact for the case of V$_{SD}$~$>$~0, and beneath the edge of the source contact for the case of V$_{SD}$~$<$~0. Specifically at an x-position 1.5 nm from the Al/AlOx edge (beneath the Al), which was chosen as it was the closest position allowable by the simulation mesh, as it was assumed that the current will flow from the Al into the Si at a position close to the edge of the S/D contacts due to the higher resistance of the Si in comparison to Al. A linear interpolation/extrapolation of this data yielded all values of $\xi$ used to determine $J_{SD}$ for direct comparison to measured data.

Synopsis TCAD simulates an ideal junction between Al-Si, therefore it is expected that the simulated values of $\xi$ will be an overestimate of the true field in the devices, because non-abruptness has the effect of reducing the electric field at a tunnel junction~\cite{Zhang}, and non-abruptness is expected for Al-Si interfaces~\cite{Brillson}. This reduction in $\xi$ will lead to a commensurate reduction in current. In order to account for the effects of non-abruptness, Eq.~\ref{eq:J} was used to determine the current density of Device 1 for a Schottky barrier height for holes of 0.35~eV~\cite{Card} for both the ideal case and with a series of constant offsets subtracted from $\xi$ in the range 0.1~MV/cm to 1.0~MV/cm. This is shown in Fig.~\ref{fg:MOD}~b). It can be seen how adjusting a constant range of $\xi$ in magnitude, greatly affects the slope of the device current's response to V$_{SD}$.

In order to make a direct comparison with the measured data of Device 1, knowing the true tunnel current area is difficult, due to uncertainty about local field disorder at the edges of the device contacts. However, upper and lower limits of the tunnel current area can be estimated, using the device channel area (2 x 245 nm) for the upper limit and assuming that tunneling only occurs at a point in each corner of a contact (1 x 1 nm) for the lower limit. The results for Device 1 taken at 50~K are included in Fig.~\ref{fg:MOD}~b) for comparison with the simulated data. It can be observed that it follows the predicted trend for a certain range of $\xi$ offsets; around a 0.4~-~0.5 MV/cm offset for the source, and around a 0.7~-~0.8 MV/cm offset for the drain; values which may account for non-abruptness at the Al-Si device contacts. Furthermore, by using these offsets in conjunction with a range of barrier heights and Eq. 2, and by making comparison to the same experimental data of Fig.~\ref{fg:MOD}~b), it was possible to confirm that the barrier height falls within the range of 0.3 - 0.4 eV for this device. Characterizing abruptness at the Al-Si interface is of importance for emerging superconducting qubit technology~\cite{Zeng}.

The presence of non-abrupt contacts is supported by the high sub-threshold swing of approximately 300~mV/dec for Device 1. One way to optimize this design and to obtain more atomically abrupt source/drain contacts would be by including a silicide layer in between the Al and Si, which would greatly improve sub-threshold swing and also enable a lower barrier height, giving better on/off current ratio~\cite{Larson}, but still allow the possibility of Al$_{y}$O$_{x}$ growth for electrical isolation from the gate. Furthermore, by slightly modifying the lithographic design, we will be able to significantly reduce the overlap between the top gate and the metal electrodes ($A$~=~$L_{overlap}$~$\ast$~$W_{device}$) to an easy achievable $A$~=~50~nm~$\ast$~50 nm~\cite{Angus, Jove}. A crude estimation of the circuit delay time~\cite{Nik}, $t$, for this optimized scenario, is in the low $ns$ range~\cite{number}, in line to what is found in the literature for T-FETs~\cite{Larson, Connelly, Nik}.

In conclusion, we have presented an SB-MOSFET compatible with CMOS and superconducting technology operated at cryogenic temperatures. Furthermore, we have demonstrated the operation of an SB-MOSFET with the gate overlapping the source/drain electrodes, and working at temperatures as low as 10~mK. We have provided simulation of the device electric field and shown that tunnel current can be predicted by Eq.~\ref{eq:J} for SB-MOSFETs and give an indication of the effects of Al-Si interface abruptness. It is the belief of the authors that the basis for this device structure holds strong potential for on-chip integration with nano-scale cryogenic Si devices.

The Authors would like to acknowledge J. Muhonen, J. Salfi and M. Veldhorst for helpful insight and discussion. We acknowledge the support of the Australian National Fabrication Facility for device manufacturing. T. L. D. and S.R. acknowledge the ARC FT scheme, project ID's: FT100100025 and FT100100589, respectively, and G. C. Tettamanzi acknowledges the ARC-DECRA scheme, project ID: DE120100702.


\begin{thebibliography}{99}
\bibitem{number} $t$~$\approx$~$V_{dd}$~$\ast$~$C_{gsd}$~$\ast$~$W_{device}$~/~$I_{dd}$, with $V_{dd}$~=~0.5~V, $I_{dd}$~=~15~nA, and $C_{gsd}$~=~$\epsilon_{AlyOx}$~$\ast$~$\epsilon_{0}$~$\ast$~$L_{overlap}$~/~$T_{AlyOx}$, where $\epsilon_{AlyOx}$ is assumed to be ~10, $\epsilon_{0}$~$\approx$~1E-11~F/m and $T_{AlyOx}$, the oxide thickness, is assumed to be $\approx$~3-5~nm.
\end{thebibliography}
\end{document}